\documentclass[manuscript,nonacm]{acmart} %
\setcopyright{none} %

\AtBeginDocument{%
  }

\setcopyright{acmlicensed}
\copyrightyear{2025}
\acmYear{2025}
\acmDOI{XXXXXXX.XXXXXXX}

\acmConference[Conference acronym 'XX]{Make sure to enter the correct
  conference title from your rights confirmation emai}{June 03--05,
  2018}{Woodstock, NY}
\acmISBN{978-1-4503-XXXX-X/18/06}

\PassOptionsToPackage{dvipsnames,table}{xcolor}
\PassOptionsToPackage{switch}{lineno}
\usepackage{multicol}
\usepackage{tabularx}
\usepackage[frozencache,cachedir=minted-cache]{minted}
\usepackage{graphicx} %
\usepackage{subcaption} %
\usepackage[textsize=tiny,disable]{todonotes}  %
\presetkeys{todonotes}{fancyline}{}
\usepackage{xspace}
\usepackage{relsize}
\usepackage[inline]{enumitem} %
\usepackage{tikz}
\usepackage{makecell} %
\usepackage{qtree}
\usepackage{threeparttable}
\usepackage[table]{xcolor}
\usepackage{cleveref}
\usepackage{colortbl}
\usepackage{algorithm}
\usepackage[noend]{algpseudocode}
\usepackage{amsthm}
\usepackage{wrapfig}

\newcommand{\circled}[1]{\tikz[baseline=(char.base)]{
    \node[shape=circle,draw,inner sep=1pt, fill=gray!30] (char) {#1};}}

\algnewcommand\algorithmicswitch{\textbf{switch}}
\algnewcommand\algorithmiccase{\textbf{case}}
\algnewcommand\algorithmicassert{\texttt{assert}}
\algnewcommand\Assert[1]{\State \algorithmicassert(#1)}%
\algdef{SE}[SWITCH]{Switch}{EndSwitch}[1]{\algorithmicswitch\ #1}{\algorithmicend\ \algorithmicswitch}%
\algdef{SE}[CASE]{Case}{EndCase}[1]{\algorithmiccase\ #1:}{\algorithmicend\ \algorithmiccase}%
\algtext*{EndSwitch}%
\algtext*{EndCase}%
\algnewcommand{\LineComment}[1]{\State #1}

\newtheoremstyle{custom}
    {}                %
    {}                %
    {\normalfont}     %
    {}                %
    {\bfseries}       %
    {.}               %
    { }               %
    {\thmname{#1}}  %

\theoremstyle{custom}
\newtheorem{definition}{Definition}

\usetikzlibrary{decorations.pathmorphing, tikzmark}

\newcommand{\APPROACH}{{\smaller STALAGMITE}\xspace}

\newcommand{\Arvada}{Arvada\xspace}
\newcommand{\TreeVada}{TreeVada\xspace}
\newcommand{\Kedavra}{Kedavra\xspace}
\newcommand{\Mimid}{Mimid\xspace}
\newcommand{\Autogram}{Autogram\xspace}

\newcommand{\tinyc}{{\smaller TINY-C}\xspace}
\newcommand{\mjs}{{\smaller MJS}\xspace}
\newcommand{\lisp}{{\smaller LISP}\xspace}
\newcommand{\harrydcjson}{{\smaller HARRYDC-JSON}\xspace}
\newcommand{\parson}{{\smaller PARSON}\xspace}
\newcommand{\calccpp}{{\smaller CALCCPP}\xspace}
\newcommand{\cjson}{{\smaller CJSON}\xspace}
\newcommand{\calc}{{\smaller CALC}\xspace}
\newcommand{\cgidecode}{{\smaller CGI-DECODE}\xspace}
\newcommand{\json}{{\smaller JSON}\xspace}
\newcommand{\llm}{{\smaller LLM}\xspace}
\newcommand{\klee}{{\smaller KLEE}\xspace}
\newcommand{\llvm}{{\smaller LLVM}\xspace}

\definecolor{sage}{HTML}{CCD3CA}
\definecolor{beige}{HTML}{F5E8DD}
\definecolor{peach}{HTML}{EED3D9}
\definecolor{babyblueeyes}{rgb}{0.63, 0.79, 0.95}

\definecolor{termcolor}{HTML}{bb8844}
\definecolor{nontermcolor}{HTML}{009999}

\newcommand{\conclusion}[1]{%
  \begin{center}
    \fbox{\begin{minipage}{0.95\columnwidth}
      \centering
      \emph{#1}
    \end{minipage}}
  \end{center}
}

\def\term#1{\textbf{#1}}
\def\nonterm#1{\textnormal{\emph{#1}}}

\begin{document}

\title{Inferring Input Grammars from Code with~Symbolic~Parsing}

\author{Leon Bettscheider}
\affiliation{%
  \institution{CISPA Helmholtz Center for Information Security}
  \city{Saarbrücken}
  \country{Germany}}
\email{leon.bettscheider@cispa.de}

\author{Andreas Zeller}
\affiliation{%
  \institution{CISPA Helmholtz Center for Information Security}
  \city{Saarbrücken}
  \country{Germany}}
\email{zeller@cispa.de}

\renewcommand{\shortauthors}{Bettscheider and Zeller}

\begin{abstract}
  Generating effective test inputs for a software system requires that these inputs be \emph{valid,} as they will otherwise be rejected without reaching actual functionality.
  In the absence of a specification for the input language, common test generation techniques rely on \emph{sample inputs}, which are abstracted into matching grammars and/or evolved guided by test coverage.
  However, if sample inputs \emph{miss} features of the input language, the chances of generating these features randomly are slim.

  In this work, we present the first technique for \emph{symbolically} and \emph{automatically} mining input grammars from the code of recursive descent parsers.
  So far, the complexity of parsers has made such a symbolic analysis challenging to impossible.
  Our realization of the \emph{symbolic parsing} technique overcomes these challenges by
  \begin{enumerate*}[label=(\arabic*)]
    \item associating each parser function~\texttt{parse\_ELEM()} with a nonterminal~\texttt{<ELEM>};
    \item limiting recursive calls and loop iterations, such that a symbolic analysis of~\texttt{parse\_ELEM()} needs to consider only a finite number of paths; and
    \item for each path, create an expansion alternative for~\texttt{<ELEM>}.
  \end{enumerate*}
  Being purely static, symbolic parsing does not require seed inputs; as it mitigates path explosion, it scales to complex parsers.

  Our evaluation promises symbolic parsing to be highly accurate.
  Applied on parsers for complex languages such as \tinyc or \json,
  our \APPROACH{} implementation extracts grammars with an accuracy of 99--100\%, widely improving over the state of the art despite requiring only the program code and no input samples.
  The resulting grammars cover the entire input space, allowing for comprehensive and effective test generation, reverse engineering, and documentation.
  \end{abstract}


\maketitle

{
\small\linespread{1.05}
\begin{figure}[t]
\begin{center}
\small
\begin{minted}[escapeinside=!!]{xml}
<start> ::= <json_parse_value>
<json_parse_value> ::=
      <opt_sign> <number>
    | <opt_sign> <nan>
    | <opt_sign> <inf>
    | '[' <json_parse_array>
    | '{' <json_parse_object>
    | '"' <opt_any_str> '"'
    | 'null' | 'false' | 'true'
<opt_any_str> ::= '' | <any_char> '+'
<opt_sign> ::= '' | '+' | '-'
<inf> ::= /[iI][nN][fF]/ | /[iI][nN][fF][iI][nN][iI][tT][yY]/
<nan> ::= /[nN][aA][nN]/
<number> ::= <main> <opt_exponent>
<json_parse_array> ::= <json_parse_array!$'$!> | ']'
<json_parse_object> ::= <json_parse_object!$'$!> | '}'
<any_char> ::= /[\x01-\xff]/
<main> ::= <digits> | <opt_digits> '.' <digits>
<opt_exponent> ::= '' | '[eE]' <opt_sign> <digits>
<digits> ::= <digit>+
<opt_digits> ::= '' | <digits>
<digit> ::= /[0-9]/
<json_parse_array!$'$!> ::= <json_parse_value> ',' <json_parse_array!$'$!>
    | <json_parse_value> ']'
<json_parse_object!$'$!> ::=
      <json_parse_value> ':' <json_parse_value> ',' <json_parse_object!$'$!>
    | <json_parse_value> ':' <json_parse_value> '}'
    | '}'
\end{minted}
\end{center}
\caption{\harrydcjson grammar mined by \APPROACH}%
\label{fig:json-grammar}
\end{figure}
}

\section{Introduction}
\label{sec:introduction}

Generating test inputs for computer programs continues to be one of the big challenges of software testing.
The biggest challenge is that one needs inputs that are \emph{valid} (such that they are not rejected by the parser) and at the same time \emph{uncommon} (such that one can test corner cases not found in common inputs).
Totally \emph{random} inputs, for instance, would be uncommon, but also very likely invalid.
\emph{Sample} inputs collected or condensed from public sources (say, via an \llm), on the other hand, tend to be \emph{valid}, but also \emph{common.}

In order to obtain inputs that are valid \emph{and} uncommon, the textbook solution is to \emph{specify} the input language.
Using a formal \emph{grammar}, we can use the grammar as a \emph{producer} and thus generate inputs that are syntactically \emph{valid} in the first place.
\Cref{fig:json-grammar}, for instance, shows a grammar for the \harrydcjson{} parser.
Having this grammar has several benefits:
\begin{itemize}
  \item Every input produced by this grammar is \emph{valid,} i.e., accepted by \harrydcjson.
  Using this grammar and a fast producer~\cite{gopinath2019buildingfastfuzzers}, we can easily produce millions of valid test inputs within minutes.
  \item The grammar covers the entire input language of \harrydcjson, including every feature (including \emph{uncommon} ones); using it as a producer will test every \harrydcjson feature.
  \item We can use the grammar to detect \emph{deviations} from standards.
  The string \texttt{\char123 true: infinity,\char125} is accepted by the grammar in \Cref{fig:json-grammar} (and \harrydcjson), but in multiple ways invalid according to the \json standard.\footnote{In the \json standard,
  \begin{enumerate*}[label=(\arabic*)]
    \item object keys (\texttt{true}) must be strings;
    \item \texttt{infinity} is not a legal value; and
    \item trailing commas are not allowed.
  \end{enumerate*}
  }
  \item We can use it to \emph{parse} existing inputs, check their validity or extract input fragments.
  \item We can \emph{mutate} existing inputs while preserving validity---all this with much higher efficiency than with random inputs or random lexical mutations, and with much higher language coverage than sample inputs.
\end{itemize}
All these benefits of grammars are well recognized in the testing community.
The catch, however, is the widespread absence of formal specifications, including grammars.
It thus is no surprise that recent research has focused on successfully \emph{mining grammars} from code and/or input samples:
\begin{itemize}
  \item \emph{\Autogram}~\cite{hoeschele2016autogram} and \emph{\Mimid}~\cite{gopinath2020mimid} introduced the concept of \emph{dynamic grammar mining.} Given a program~$P$ and a set of sample inputs, they dynamically track how~$P$ processes input characters, and form tokens and grammar structures from characters processed similarly.
  \item \emph{\Arvada}~\cite{kulkarni2021arvada} learns grammars from a set of sample inputs, using a \emph{parse oracle} to check which input parts can be replaced by others (and thus form a grammar abstraction).
\end{itemize}
The downside of these approaches, however, is that they are \emph{biased} towards the set of sample inputs; if a language feature is not present in the samples already (like \texttt{NaN} in \json samples), they cannot observe it and thus not incorporate it into the mined grammar.
And of course, if no sample inputs are available, these approaches become entirely inapplicable, as there are neither executions to observe nor samples to mine grammars from.

How can we avoid this reliance on sample inputs? 
So far, the problem has been that \emph{symbolic} approaches (which need no inputs) are effective in the small, but always have \emph{failed} to analyze nontrivial parsers---the high number of branches, loops, and recursive calls, all intertwined, results in a combinatorial explosion of paths that was impossible to manage.

\begin{figure}[t] %
  \centering
  \includegraphics[width=\textwidth]{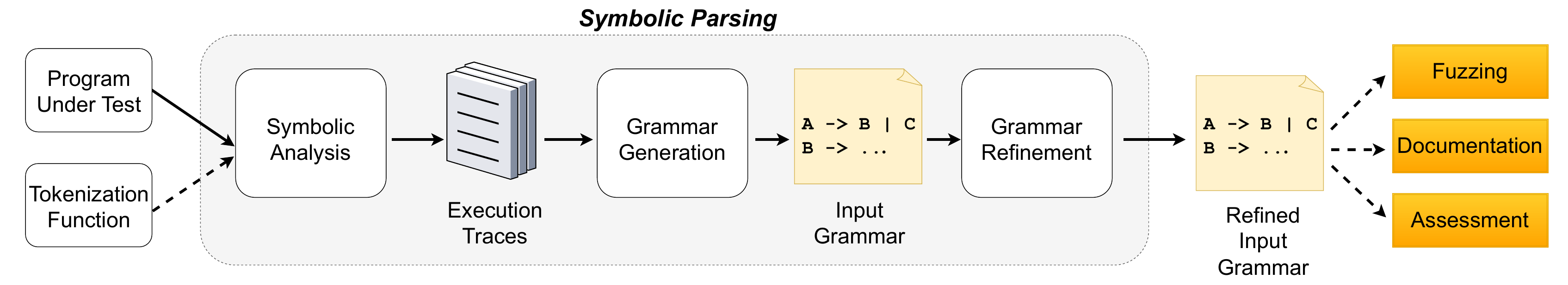}
  \caption{
    How \APPROACH works.
    \APPROACH infers context-free input grammars from recursive-descent parsers.
    It starts by symbolically executing and tracing the program under test, comprehensively exploring execution paths by limiting loop iterations and recursive calls, while tracking \emph{input consumptions} by the parser.
    Subsequently, these symbolic execution traces are converted into an input grammar by leveraging execution context information.
    Finally, this input grammar is refined to reduce overapproximation.}
  \label{fig:overview}
\end{figure}

In this paper, we introduce \APPROACH\footnote{\APPROACH stands for \underline{sta}tic \underline{l}anguage \underline{a}nd \underline{g}rammar \underline{mi}ning for \underline{te}sting.}, the first scalable \emph{fully automatic static grammar miner.}
From a program under test with an input parser (\Cref{fig:overview}), \APPROACH automatically extracts the input language as a grammar.
The \json input grammar in~\Cref{fig:json-grammar}, for instance, is automatically mined from \harrydcjson by \APPROACH.

This grammar
\begin{itemize}
  \item is extracted from code alone, i.e.\ not requiring any input samples;
  \item has a \emph{recall} of 100\% (i.e., it covers the entire input language);
  \item has a \emph{precision} of 100\% (i.e., all strings produced are valid);
  \item can easily serve as \emph{producer,} \emph{parser,} and \emph{mutator} for efficient fuzzing; and
  \item is sufficiently structured and readable for reverse engineering and documentation.
\end{itemize}

\APPROACH takes as input a C program with a \emph{recursive descent} input parser and produces its input grammar.\footnote{The alternative \emph{table-driven parsers} are generated from a grammar specification; for these, a formal grammar is there in the first place and does not need to be extracted.
Recently, the CLANG and GCC compilers switched from table-driven parsers to recursive descent parsers to gain larger control over the parsing process.}
If the program under test uses a separate \emph{lexer} to process characters into tokens, an additional \emph{harness} of 5--10 lines is required to specify the information flow between lexer and parser. Other than that, \APPROACH operates fully automatically and extracts the complete grammar within minutes to hours.

To realize symbolic parsing (and hence static grammar mining), we tackle the path explosion problem in the context of recursive descent parsers by limiting the exploration of loops and recursion, facilitating a comprehensive symbolic exploration.
During symbolic execution, we track \emph{consumptions of input characters} and associate them to the current execution context (i.e., call path and loop iterations).
This context information allows constructing \emph{parse trees} of explored inputs.
Subsequently, we generalize these parse trees to input grammars:
\begin{itemize}
\item Inner nodes of the parse trees (i.e., functions and loops) are converted to non-terminal symbols in the grammar. This is consistent with the implementation style of recursive descent parsers, which define a function for each non-terminal symbol.
\item Bounded control structures (i.e., loop iterations and recursion) are generalized and again instantiated, allowing for unbounded repeated elements (loops) and nested elements (recursion) again.
\item Lexical elements (i.e., tokens) are generalized by matching observations with predefined string classes (such as lower case strings or digits).
\end{itemize}

Note, though, that while this paper demonstrates the feasibility of symbolic parsing in practice, it would be overly optimistic at this point to expect that one can place an arbitrary program into a black box and have the input grammar pop out.
Complex input processors with multiple stages, as found in compilers, are very much out of reach for our implementation, as are binary input formats with features such as length encodings, offsets, or checksums that cannot be represented adequately in a context-free grammar.
Hence, there are still years of research and engineering ahead until one can claim a one-stop solution---even if this work shows a new and very promising path towards such a solution.

In summary, we make the following contributions:
\begin{enumerate}
  \item We introduce the \emph{first approach to statically mine grammars from code} without requiring input samples.
  \item We \emph{detail} how we realized \emph{symbolic parsing} in our \APPROACH prototype.
  \item We \emph{evaluate} \APPROACH and show that it produces highly accurate grammars with precision and recall of 99--100\%.
\end{enumerate}

The remainder of this paper is organized as follows.
\textbf{\Cref{sec:symbolic_exploration}} introduces our symbolic execution-based technique for exploring parsers and producing execution traces.
\textbf{\Cref{sec:inferring_grammar}} explains how we infer inputs grammars from these execution traces.
\textbf{\Cref{sec:implementation}} provides implementation details of \APPROACH.
In~\textbf{\Cref{sec:evaluation}}, we evaluate \APPROACH on a set of parsers, assessing the accuracy and utility of the mined grammars.
Related work is discussed in~\textbf{\Cref{sec:related_work}}, followed by limitations and future work in~\textbf{\Cref{sec:limitations_future_work}}.
We close with our conclusion and key takeaways in \textbf{\Cref{sec:conclusion}}.
The \APPROACH tool and all experimental data are available as open source (\textbf{\Cref{sec:data-availability}}).

\section{Symbolically Exploring Parsers and Generating Execution Traces}
\label{sec:symbolic_exploration}

In this section, we present the first step of symbolic parsing: tracing and bounding the symbolic execution of parsers.
The objective of this step is to collect an execution trace for each symbolically executed control-flow path.
Each trace maps input positions to execution contexts, consisting of the call path and loop iterations that were active during consumption of this input position, as well as all possible solutions.
A sample execution trace is shown in~\Cref{tab:sample-execution-trace}.
To briefly anticipate the next step, which we detail in~\Cref{sec:inferring_grammar}, these execution traces are later converted into parse trees (\Cref{fig:parse-tree}), which are then aggregated and generalized to infer inputs grammars.

To ensure meaningful results, we only consider execution paths that lead to a successful parse.
Determining whether an input was successfully parsed is straightforward, as parsers typically return a non-zero value or throw an exception to indicate an error.
Throughout this section, we refer to~\Cref{fig:algorithm1}, showcasing how the individual components of \APPROACH may be implemented in a symbolic execution engine.

\begin{figure}[h]
  \centering
  \begin{minipage}{0.45\textwidth}
  \rowcolors{2}{gray!20}{white}
  \begin{tabular}{l l c}
      \rowcolor{gray!20}
      \textbf{\#} & \textbf{Execution Context} & \textbf{Solutions} \\
      0
      & parse:value
      & \texttt{[}
      \\

      1
      & parse:value:array:L1I1:value%
      & \texttt{/[0-9]/}
      \\

      2
      & parse:value:array:L1I1%
      & \texttt{,}
      \\

      3
      & parse:value:array:L1I2:value
      & \texttt{"}
      \\

      4
      & parse:value:array:L1I2:value%
      & \texttt{"}
      \\

      5
      & parse:value:array:L1I2%
      & \texttt{]}
      \\
  \end{tabular}
  \caption{A simplified execution trace for \harrydcjson}
  \label{tab:sample-execution-trace}
\end{minipage}%
\hfill
\begin{minipage}{0.45\textwidth}
  \Tree [.\nonterm{parse}
  [.\nonterm{value}
     {\term{[}}
      [.\nonterm{array}
        [.\nonterm{Loop 1}
          [.\nonterm{Iteration 1}
            [.\nonterm{value}
                {\term{/[0-9]/}}  
            ]
              {\term{,}}
          ]
          [.\nonterm{Iteration 2}
            [.\nonterm{value}
              {\term{"}}
                {\term{"}}  
            ]
              {\term{]}}
          ]
        ]
      ]
    ]
  ]
   \caption{Parse tree derived from~\Cref*{tab:sample-execution-trace}}
   \label{fig:parse-tree}
\end{minipage}
\vspace{-0.5cm}
\end{figure}

\subsection{Assumptions}
\label{sec:assumptions}

We start by defining recursive-descent parsers, as \APPROACH focuses on mining grammars from such parsers.%

\begin{definition}
  A recursive-descent parser consists of a set of \emph{mutually recursive parse functions.}
  Each parse function corresponds to one non-terminal symbol in the grammar and recognizes the portion of the language that this symbol defines.
  To parse a non-terminal symbol \mintinline{xml}{<B>} in a production rule of \mintinline{xml}{<A>}, the parse function \texttt{A} calls parse function \texttt{B}.
  To recognize a terminal symbol \mintinline{xml}{"b"} in a production rule of \mintinline{xml}{<A>}, function \texttt{A} compares characters from the input stream~\cite{cooper2011engineering}.
  \end{definition}

We make the following assumptions about the program under test:

\begin{itemize}
  \item The program under test is a \emph{recursive-descent parser}.
  \item The program under test incrementally processes and accepts characters from the input stream.
  \item The program under test only checks syntactic input features that can be expressed in a context-free grammar.
  \item The program under test can be compiled to \llvm IR (e.g., C and C++ programs).
\end{itemize}

  \subsection{Tracking Input Consumption}
  \label{sec:tracking_input_consumption}

One of the key challenges in symbolic grammar inference is identifying the exact point in execution when the parser consumes each input character.
This is essential because we utilize the execution context, which is active during the consumption of each input character, to construct the input grammar.
Specifically, we leverage the hierarchical information provided by the currently active call path and loop iterations.

\begin{definition}
  We define the \emph{input consumption} of an input character as the point in the execution at which the parser recognizes the character and no longer needs to process it again for parsing purposes.
\end{definition}

Our method for tracking input consumptions is similar to the one used by Mimid~\cite{gopinath2020mimid}, but it differs in two significant ways.
Before outlining these differences, we first provide an overview of the technique used by Mimid.

\begin{description}[]
\item[The \Mimid Approach]
Mimid is a dynamic input grammar mining technique that uses lightweight instrumentation to track how a parser consumes input characters of a given concrete input.
Specifically, it monitors accesses only to the original input buffer, assuming that the parser neither modifies nor duplicates it.
Since a parser may access a character multiple times before ultimately consuming it, Mimid attributes each character to the execution context of its last recorded access.

\item[The \APPROACH Approach]
\mbox{\APPROACH} differs from \Mimid in two key ways when tracking input consumption.
First, instead of monitoring only the original input buffer, we track accesses to all buffers that contain input characters using a lightweight data-flow tracking approach.
Second, we do not only record the \emph{last} access but \emph{all} accesses to input characters.
Subsequently, as we discuss in~\Cref{sec:identifying_input_consumptions}, we apply a heuristic to find the input accesses that actually correspond to input consumptions.
In the following section, we detail these two aspects of our approach.
\end{description}

\subsubsection{Lightweight Data-Flow Tracking}
\label{sec:lightweight_data_flow_tracking}

We observed that some parsers may copy portions of the input buffer into an other buffer during parsing.
As a result, tracking accesses only to the original input buffer can lead to inaccuracies.
For instance, consider the code snippet in~\Cref{fig:cjson-code} from one of our evaluation subjects, \cjson, where a substring of the input buffer is copied into an other buffer~\circled{1} and subsequently processed further~\circled{2}.

\begin{figure}[b]%
  \centering
  \small
  \begin{minted}[linenos,escapeinside=||]{C}
static cJSON_bool parse_number(cJSON * const item,
                parse_buffer * const input_buffer) {
    /* ... */
    for (i = 0; /* ... */; i++) {
        switch (buffer_at_offset(input_buffer)[i])
        {
            case '0': case '1': case '2': case '3': case '4':
            case '5': case '6': case '7': case '8': case '9':
            case '+': case '-': case 'e': case 'E':
                number_c_string[i] = buffer_at_offset(input_buffer)[i]; |\circled{1}|
                break;
            /* ... */
            default: goto loop_end;
        }
    }
loop_end:
    number_c_string[i] = '\0';
    number = strtod((const char*)number_c_string, (char**)&after_end); |\circled{2}|
    /* ... */
}
  \end{minted}
  \vspace{-.4cm}
  \caption{\cjson copies a part of the input buffer to the buffer \emph{number\_c\_string} during parsing.}
  \label{fig:cjson-code}
\end{figure}

We address this issue using a lightweight data-flow tracking approach.
Specifically, we \emph{track accesses into direct copies of the input buffer}, but \emph{we do not track accesses into buffers that contain \emph{transformed} input characters}, such as \texttt{(input[0] - 0x30)} or \texttt{(input[0] + input[2])}.
The latter would indicate an evaluation performed by parsers that do not strictly separate parsing from interpretation, which we deliberately avoid tracking.
In contrast, parsing code is expected to process unmodified input characters (but may very well process copies of them).
For example, we treat the buffer \texttt{number\_c\_string} in~\Cref{fig:cjson-code} as a \emph{copy} of the input buffer, and track accesses to it.
This is straightforward to implement in a symbolic execution based technique like \APPROACH, as the symbolic store already keeps track of the symbolic input.
More precisely, this is implemented in Lines 18---19 of~\Cref{fig:algorithm1}.

\subsubsection{Recording All Input Accesses}
\label{sec:recording_all_input_accesses}
We record all accesses to the input buffer, labeling each access with an incremental number that we call the \emph{access order}, and store this along with the execution context in the trace (Lines 20-21 of~\Cref{fig:algorithm1}).
Once the parser under test has terminated successfully, we query the solver of the symbolic execution engine for all possible solutions for each input character, and finally output the trace.
This is implemented in Lines 23---25 of~\Cref{fig:algorithm1}.

\begin{algorithm}[b]%
  \caption{\APPROACH core modifications to interpretation-based symbolic execution engines}
  \label{fig:algorithm1}
    \begin{algorithmic}[1]
      \Function{symbolicInterpreter}{ExecutionState es}
          \While{running}
            \Switch{event}
             \LineComment{\ldots}
              \Case{\emph{Call from call site cs to function f}} \Comment{Limiting Recursion}
              \State es.stackFrames.push((cs, f, loopIterationStack=stack()))
              \LineComment{// Analogously, on return: pop from es.stackFrames}
              \If{es.stackFrames.count((cs, f)) > 3}
                  \State \textsc{terminateState}(es)
              \EndIf
              \LineComment{\ldots}
              \EndCase
              \Case{\emph{Enter loop header h}} \Comment{Limiting Loops}
                \State es.stackFrames.last().loopIterationStack.push(h)
                \LineComment{// Analogously, on loop exit: pop from loopIterationStack}
                \If{es.stackFrames.last().loopIterationStack.count(h) > 4}
                  \State \textsc{terminateState}(es)
                \EndIf
                \LineComment{\ldots}
              \EndCase
              \Case{\emph{Load from address a}} \Comment{Tracking Input Consumption}
                \State symExpr $\gets$ load(a)
                \If {symExpr == (Read symbolicInputBuffer inpPos)}
                  \State es.trace[inpPos]['accessOrders'].push(es.accessOrder++)
                  \State es.trace[inpPos]['executionContext'].push(es.stackFrames.serialize())
                \EndIf
                \LineComment{\ldots}
              \EndCase
            \EndSwitch
          \EndWhile
  
          \For{inpPos \textbf{ in } es.trace.keys()}
            \State es.trace[inpPos]['solutions'] $\gets$ \textsc{solve}(es, (Read symbolicInputBuffer inpPos))
          \EndFor
          \State \textsc{serializeTrace}(es.trace)
      \EndFunction
    \end{algorithmic}
  \end{algorithm}

\subsection{Limiting Loops}
\label{sec:limiting_loop_iterations}
Next, we describe our technique to bound loops in parsers.
The purpose is to enable comprehensive symbolic exploration by mitigating path explosion.
Before symbolic execution starts, we retrieve the header, exiting, and exit basic blocks of each loop in the program.
During symbolic execution, we first mark loops as \emph{input-processing} if they access multiple input characters.
Loops are assigned the \emph{input-processing} attribute only for the current execution context, allowing loops to take different roles, e.g., for different calling contexts.
We effectively limit the exploration of \emph{input-processing} loops to at most four iterations.
This number was chosen based on our observations of the parsers we evaluated.
We do so by counting control flow transitions to the header basic block of each loop, and terminate the current execution state when a loop would be entered for the fourth time.
This is implemented in Lines 11---15 of~\Cref{fig:algorithm1}.

\subsection{Limiting Recursion}
\label{sec:limiting_recursion}
Similar to how we limit loops, we also bound recursive calls to a recursion depth of three.
We track the call path during symbolic execution as a vector of \emph{(caller, callee)} pairs within the symbolic execution state.
When a function is called, we check if the current \emph{(caller, callee)} pair appears more than three time in the call path.
If it does, we terminate the current execution state, effectively bounding recursive calls.
Specifically, this is implemented in Lines 5---9 of~\Cref{fig:algorithm1}.
The number \emph{three} was chosen based on our observations of the parsers
we evaluated.
In principle, this number could be increased if necessary, which would also increase the number of paths explored and thus the complexity.

\subsection{Handling Parsers With a Lexing Stage}
\label{sec:handling-lexers}

In ad-hoc tokenization, which is assumed in \Cref{sec:tracking_input_consumption}, parse functions only check for one specific token at a time.
However, some parser implementations follow a different methodology and contain a single dedicated tokenization function, which is called by all parse functions and which may read \emph{any} of the possible tokens.
Symbolic execution of such parsers is challenging~\cite{pan2021grammar} because the tokenization function is called multiple times during symbolic execution and often contains many branches and loops.
We found that this circumstance routinely overburdens symbolic execution.

To address this problem, \APPROACH analyzes the tokenization function only once in isolation, storing pairs of \emph{token instance} (e.g., "123") and associated \emph{token identifier} (e.g., a numerical value representing an \texttt{INT} token) that was returned by the tokenization function, in a token grammar.
When this tokenization function is called during the symbolic execution of the parser, we detour this call to a \emph{proxy function} that simply returns a \emph{symbolic token identifier} without processing the input.
This drastically reduces the burden on the symbolic execution engine.

Some manual work is required to implement
\begin{enumerate*}[label=(\arabic*)]
  \item the \emph{harness} to symbolically execute the tokenization function \emph{in isolation}; and
  \item the \emph{proxy function,} which serves as \emph{glue code} when analyzing the whole parser.
\end{enumerate*}
The harness needs to mark the input variable symbolic, call the tokenization function, check parsing success, and pass the token identifier to the symbolic execution engine; this work is easy to do for a developer, requiring less than 10 lines of code.
Automating this task is difficult, however, as the interface between the parser and the tokenization function is not standardized.
In fact, tokenization functions may return the token identifier in many different ways, such as via a global variable, a return value, or an output parameter.
Similarly, the tokenization function may read the input in different ways.

\Cref{fig:harness-tokenization-function} shows the harness for the tokenization function \emph{next\_sym()} of \tinyc, one of our evaluation subjects.
It sets up \emph{inp} as a symbolic input, and writes it to the global \emph{cursor} variable used by \emph{next\_sym()}, which is then called.
Finally, it passes the token identifier \emph{sym} to the symbolic execution engine, which logs it as part of the trace.
\Cref{fig:proxy-tokenization-function} shows the proxy function for \emph{next\_sym()}, which is used instead during parser analysis.
It simply assigns a symbolic token identifier to the global variable \emph{sym} holding the current token identifier.

For parsers with a lexing stage, we use the same technique to track input consumptions as we do for ad-hoc tokenization, except that we record accesses to symbolic token identifiers instead of input characters.

\begin{figure}[h]
  \centering
  \begin{minipage}[t]{0.45\textwidth}
  \begin{minted}{C}
void wrapper_next_sym() {
  char* inp = stalagmite_sym_input();
  cursor = &inp;
  next_sym();
  stalagmite_solve_and_outp_token(sym);
}
  \end{minted}
  \caption{Harness for \tinyc tokenizer \emph{next\_sym()}}
  \label{fig:harness-tokenization-function}
  \end{minipage}%
  \hfill %
  \begin{minipage}[t]{0.45\textwidth}
  \begin{minted}{C}
void proxy_next_sym() {
  sym = stalagmite_next_sym_token();
}



  \end{minted}
  \caption{Proxy tokenization function for \mbox{\tinyc}}
  \label{fig:proxy-tokenization-function}
  \end{minipage}
\end{figure}

\section{Inferring the Input Grammar From Execution Traces}
\label{sec:inferring_grammar}

At this point, symbolic execution has completed, producing multiple execution traces (e.g., \Cref{tab:sample-execution-trace}), each representing a distinct control-flow path.
Additionally, for parsers with a lexing stage, the tokenization function analysis has generated pairs of \emph{token identifiers} and \emph{token instances}.
Next, to infer the input grammar, we proceed with the two steps briefly outlined below, which we discuss in more detail in the following sections.

\begin{description}
  \item[Converting Traces into a Grammar.] We first convert execution traces into parse trees, leveraging the execution context of input consumptions to label tree nodes.
  Subsequently, we merge these parse trees and convert them into a single grammar.
  Finally, we reinstantiate and generalize loops in this grammar.
  \item[Generalizing Tokens in the Grammar.]
  We process each token identifier and associated token instances.
  If all token instances are identical, the token is likely a \emph{keyword} or \emph{operator}. Hence, we do not process it any further.
  However, if many different token instances are associated to the same token identifier, we generalize them into one of our predefined string sets that fits best.
\end{description}

\subsection{Converting Traces to an Input Grammar}
\label{sec:syntax_analysis}

Constructing the input grammar requires the execution contexts of input consumptions.
Up to this point, we have recorded all accesses to input characters, and we now identify which of these are input consumptions.

\subsubsection{Identifying Input Consumptions in Input Accesses}
\label{sec:identifying_input_consumptions}

\begin{wrapfigure}{r}{0.4\textwidth}
  \begin{center}
  \begin{minipage}{0.39\textwidth}
  \small
  \begin{minted}[escapeinside=||]{C}
static int parse_sum(const char **pexp,
  char expect, int *value) {
  int rh, ret;
  ret = parse_mult(pexp, expect, value);
  if (!ret)
    while (**pexp != expect) {
      char op = **pexp; |\circled{1}|

      ret = ERR_TOK;
      if (op != '+' && op != '-') |\circled{2}|
        break;
      (*pexp)++;  |\circled{3}|
      ret = parse_mult(pexp, expect, &rh);
      if (ret)
        break;

      switch (op) { |\circled{4}|
      case '+':
        *value += rh;
        break;
      case '-':
        *value -= rh;
        break;
      } }
  return ret;
}
  \end{minted}
  \end{minipage}
  \caption{This \calc function combines parsing and evaluation.}
  \label{fig:calc-code}
  \end{center}
\end{wrapfigure}

In most cases, it is reasonable to assume that the last access to an input character corresponds to its point of consumption, following the approach of Mimid.
However, we observed instances where the last access does not align with the actual point of consumption.
For example, consider the function in~\Cref{fig:calc-code} from our \calc evaluation subject.

This function handles the addition and subtraction of two expressions.
From an input consumption tracking perspective, the key detail is that the current input character is first copied into the buffer \texttt{op} \circled{1}.
It is then consumed \circled{2}, assuming it is either \texttt{"+"} or \texttt{"-"} (otherwise, the parse would be invalid).
The actual consumption is evident when the input pointer is incremented \circled{3}.
However, the same input character is accessed again \circled{4} because \calc evaluates the arithmetic expression as part of parsing.
As a result, relying on the last access as the point of consumption would produce incorrect results.

To address this issue, we propose a general heuristic to identify input consumptions, aiming to map each input character to its correct point of consumption.
First, as detailed in~\Cref{sec:recording_all_input_accesses}, we record \emph{all accesses to each input character} during symbolic execution, including their execution context, and assign incremental labels to track their order.
Next, we apply our heuristic on these recorded accesses to determine the point of consumption for each input character.

\begin{figure}[t]
      \centering
      \begin{minipage}{.8\textwidth}
      \includegraphics[width=\linewidth]{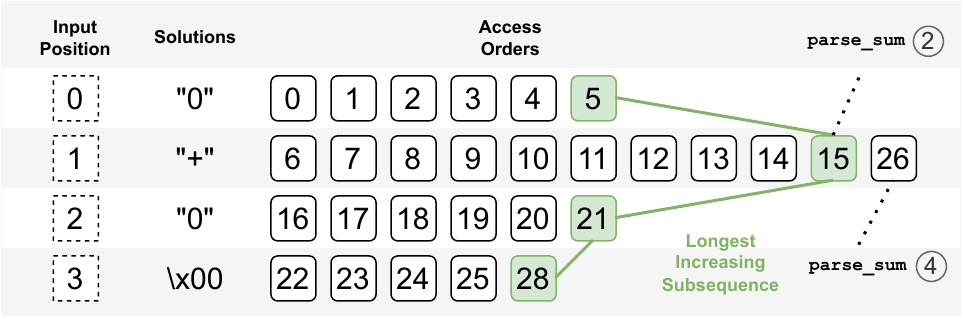}
      \end{minipage}
      \caption{A sample execution trace of the \calc subject, illustrating how input consumptions are identified. Execution contexts are omitted for brevity.}
      \label{fig:identify-consumptions}
\end{figure}

As an example,~\Cref{fig:identify-consumptions} shows an execution trace from the \calc subject for the input \texttt{0+0}.
Notably, the character at position 1 was subsequently accessed ten times (labeled 6 to 15).
However, before its final access (labeled 26), the characters at position 2 and 3 were accessed.
Simply using the last accesses as the point of consumption (i.e., at labels 5, 26, 21, 28) would lead to an incorrect result.
Specifically, the access labeled 26 would cause the "+" character to be mapped to \circled{4} instead of \circled{2} in~\Cref{fig:calc-code}.

To resolve this issue, we apply~\Cref{fig:algorithm-identifying} to identify input consumptions.
The goal of this algorithm is to select and output one \emph{access order} for each input character such that the selected access orders are non-decreasing.
This algorithm starts by initializing the \texttt{resultAccessOrders} list with the last access order of each input character (Line 10), which, for the running example, is \texttt{[5,26,21,28]}.
It then calculates the longest increasing subsequence (LIS) of these access orders, which, in the running example, is \texttt{[5,21,28]}.
Since the LIS is shorter than the number of input characters (Line 12), the algorithm identifies the first outlier --- the access order corresponding to the first input character not part of the LIS --- and backtracks to the previous access of that character (Line 15).
In the example, the algorithm identifies \texttt{26} as the outlier and pops it from the \texttt{traces[1]['accessOrders']} list of the second input character.
In the next iteration, the algorithm computes the LIS of \texttt{[5,15,21,28]}, finds that its length matches the number of input characters, and thus returns \emph{resultAccessOrders}.
Until now, we have not discussed Line 5---9 of~\Cref{fig:algorithm-identifying}.
This part of the algorithm serves as a fallback to handle situations where backtracking on individual character accesses does not yield a maximum increasing subsequence.
In such cases, the algorithm assigns the access orders of the previous input character to the current input character.
This approach guarantees an increase in the length of the LIS (since we allow duplicates), facilitating convergence towards a solution.

\begin{algorithm}[h]
  \caption{Identifying Input Consumptions}
  \label{fig:algorithm-identifying}
  \begin{algorithmic}[1]  
  \Function{identifyInputConsumptions}{traces}
    \While{True}
        \State $resultAccessOrders \gets []$
        \For{$inpPos \gets 0$ \textbf{to} $\text{length}(traces) - 1$}
            \If{$traces[inpPos]['accessOrders'] = []$} \Comment{Backtracking was not successful}
                \If{$inpPos = 0$}
                    \State $traces[inpPos]['accessOrders'] \gets [0]$
                \Else
                    \State $traces[inpPos]['accessOrders'] \gets traces[inpPos-1]['accessOrders']$

                \EndIf
            \EndIf
            \State $resultAccessOrders.\text{append}(traces[inpPos]['accessOrders'][-1])$
        \EndFor
    
        \State $lis \gets \text{longestIncreasingSubsequence}(resultAccessOrders)$
    
        \If{$\text{length}(lis) = \text{length}(resultAccessOrders)$}
            \Return $resultAccessOrders$
        \EndIf
    
        \State $outlier \gets \{o \in resultAccessOrders \mid o \notin lis\}[0]$
        \State $outlierPos \gets \text{accessOrderToInpPos}(outlier)$
        \State $traces[outlierPos]['accessOrders'].\text{pop}()$ \Comment{Backtrack to previous access}
    \EndWhile
  \EndFunction
  \end{algorithmic}
\end{algorithm}

\subsubsection{Deriving the Grammar}
\label{sec:deriving-the-grammar}

\Cref{fig:algorithm2} shows our approach for deriving the input grammar from execution traces.
The first step is converting each execution trace into a parse tree.
Each execution context in the execution trace corresponds to a path in the parse tree.
The parse tree is constructed by inserting paths in the order of input indices.
The leaf node of each path are the terminal solutions.
To illustrate this, \Cref{fig:parse-tree} presents the parse tree for the execution trace shown in \Cref{tab:sample-execution-trace}.

\begin{algorithm}[h]
  \caption{Converting traces to an input grammar. Calling contexts omitted for brevity.}
  \label{fig:algorithm2}

  \begin{algorithmic}[1]

    \Function{treePath}{executionContext}
        \LineComment{// Example: This function converts "parse:value:array:L1I1" to a path of four nodes:}
        \LineComment{// InnerNode("parse")$\rightarrow$InnerNode("value")$\rightarrow$InnerNode("array")$\rightarrow$LoopNode("L1I1")}
    \EndFunction

    \Function{leafNode}{solutions}
      \LineComment{// Returns a LeafNode containing all solutions}
    \EndFunction

    \Function{treeToGrammar}{parseTree}
        \State grammar $\gets$ \textsc{Grammar}()
        \For{Node N \textbf{in} parseTree}
            \Switch{N.type}
                \Case{LeafNode}
                    \State grammar[N.name].addRule([s] for s in N.solutions)
                \EndCase
                \Case{InnerNode}
                    \State rule $\gets$ [C.name \textbf{for} C \textbf{in} N.children]
                    \State grammar[N.name].addRule(rule)
                \EndCase
                \Case{LoopNode} \Comment{Loop Generalization}
                    \State grammar["L" + N.name].addRule(["L" + N.name + "\_cont", "L" + N.name])
                    \State grammar["L" + N.name].addRule(["L" + N.name + "\_exit"])
                    
                    \State rule $\gets$ [C.name \textbf{for} C \textbf{in} N.children]
                    \If{\textsc{isLastIteration}(N)}
                        \State grammar["L" + N.name + "\_exit"].addRule(rule)
                    \Else
                        \State grammar["L" + N.name + "\_cont"].addRule(rule)
                    \EndIf
                \EndCase
            \EndSwitch
        \EndFor
        \State \Return grammar
    \EndFunction

    \Function{traceToTree}{trace}
        \State parseTree $\gets$ \textsc{Tree}()
        \State consumingAccessOrders $\gets$ \textsc{identifyInputConsumptions}(trace) \Comment{\Cref{sec:identifying_input_consumptions}}
        \For{$inpPos \gets 0$ \textbf{to} $\text{length}(traces) - 1$}
            \State consumingAccessOrder $\gets$ consumingAccessOrders[inpPos]
            \State consumingExecutionContext $\gets$ accessOrderToExecCtx(traces, consumingAccessOrder)
            \State path $\gets$ \textsc{treePath}(consumingExecutionContext)
            \State leaf $\gets$ \textsc{leafNode}(traces[inpPos]['solutions'])
            \State parseTree.insert(path + leaf)
        \EndFor
    \EndFunction

    \Function{tracesToGrammar}{traces}
        \State grammar $\gets$ \textsc{Grammar}()
        \For{trace \textbf{in} traces}
            \State grammar.union(\textsc{treeToGrammar}(\textsc{traceToTree}(trace))) \Comment{\Cref{sec:syntax_analysis}}
        \EndFor
        \State \Return \textsc{generalizeTerminals}(grammar) \Comment{\Cref{sec:generalizing_tokens}}
    \EndFunction

\end{algorithmic}
\end{algorithm}

~\\

Subsequently, all parse trees are converted into a grammar by performing the following steps:
\begin{enumerate}
\item Each node is converted into a non-terminal symbol that encodes the calling context. For instance, calls from two different call sites to the function \texttt{value} will result in the non-terminal symbols \mintinline{xml}{<value>} and \mintinline[escapeinside=!!]{xml}{<value!$'$!>}. All children of the node are added as rule alternatives to the definition of that node in the grammar.
\item Loops are generalized by introducing \emph{continuing} and \emph{exiting} iterations. The last iteration is considered \emph{exiting} (because it leaves the loop), whereas other iterations are \emph{continuing}, i.e., they jump back to the loop header.
\item Terminal symbols are generalized as described in~\Cref{sec:generalizing_tokens}.
\end{enumerate}

\begin{figure}[H]
  \small
  \centering
  \begin{minted}{xml}
          <start>       ::= <parse>
          <parse>       ::= <value>
          <value>       ::= [ <array>
          <array>       ::= <L1>
          <L1>          ::= <L1_continue> <L1>
                          | <L1_exit>
          <L1_continue> ::= /[0-9]/ ,
          <L1_exit>     ::= ""]
\end{minted}
\caption{Input grammar extracted from~\Cref*{tab:sample-execution-trace} (simplified)}
\label{fig:trace-grammar}
\end{figure}

\Cref{fig:trace-grammar} shows the grammar for the parse tree in \Cref{fig:parse-tree}.
This grammar, which was derived from a single execution trace, can already produce inputs such as \texttt{[1,5,8,""]} and \texttt{[2,""]}. 
The grammar is completed step by step by processing all execution traces, and hence, parse trees, finally resulting in the grammar shown in~\Cref{fig:json-grammar}.

\subsection{Generalizing Tokens in the Input Grammar}
\label{sec:generalizing_tokens}

The purpose of this step is to generalize tokens in the input grammar.
For parsers with a lexing stage, we mined a separate \emph{token grammar}, which maps token identifiers to associated token instances, as discussed in~\Cref{sec:handling-lexers}. In this case, we generalize token instances for each token identifier.
For parsers with ad-hoc lexing, the token grammar is embedded into the grammar mined in the previous step. In this case, we only generalize token instances associated with non-terminal symbols of external functions (such as \texttt{strtod}). The reason is that external functions may not adhere to our assumptions listed in~\Cref{sec:assumptions}. For other non-terminal symbols, the token structure is already clear from the code structure.

\begin{figure}[h]
  \centering
  \begin{minipage}[t]{0.45\textwidth}
    \small
    \begin{minted}{xml}
      <INT>  ::= "9017" | "2" | "118" | ...
      <ID>   ::= "m" | "s" | "u" | "a" | ...
      <LPAR> ::= "("
      <WHILE_SYM> ::= "while"
    \end{minted}
    \caption{Token instances aggregated by token ID}
    \label{fig:token-instances-1}
  \end{minipage}%
  \hfill
  \begin{minipage}[t]{0.45\textwidth}
    \small
    \begin{minted}{xml}
      <INT>       ::= <digit>+
      <ID>        ::= <lower_char>
      <LPAR>      ::= "("
      <WHILE_SYM> ::= "while"
    \end{minted}
    \caption{Generalized token instances of~\Cref{fig:token-instances-1}}
    \label{fig:token-instances-2}
  \end{minipage}
\end{figure}

In the following, we detail our token generalization technique, which is identical for both types of lexers.
\Cref{fig:token-instances-1} shows an excerpt of the token identifiers and their associated token instances for the \tinyc subject.
We generalize these token instances to sets of characters by matching them with predefined patterns of string classes such as digits, hexadecimal digits, floating point numbers, lowercase strings, and so on.
\Cref{tab:string-classes} shows an excerpt of these predefined string classes.
The patterns we try to match get increasingly more permissive, and we choose the least permissive set that contains all observed token instances.
Moreover, we check if tokens may start with strings of whitespaces and generalize this accordingly.
If at most three token instances are observed per token identifier, we consider this a \emph{keyword or operator token}, which we do not generalize but instead include in the grammar as-is.
For instance, in the example above, we do not generalize \mintinline{xml}{<WHILE_SYM>} (as it is a keyword) but we do generalize \mintinline{xml}{<INT>}.
The result is a \emph{generalized token grammar}, illustrated in~\Cref{fig:token-instances-2}.

\begin{table}[h]
  \caption{String classes (excerpt)}
  \label{tab:string-classes}
  \rowcolors{2}{gray!20}{white}
  \begin{tabular}{lll}
    \rowcolor{gray!20}
    \textbf{Regular Expression} & \textbf{Non-Terminal} \\
    \texttt{\textbackslash d} & \mintinline{xml}{<digit>} \\
    \texttt{[a-z]} & \mintinline{xml}{<lower_char>} \\
    \texttt{[0-9a-fA-F]} & \mintinline{xml}{<hexdigit>} \\
    \texttt{[-+]?((\textbackslash d+(\textbackslash .\textbackslash d*)?|\textbackslash .\textbackslash d+)([eE][-+]?\textbackslash d+)?)} & \mintinline{xml}{<float_simple>} \\
  \end{tabular}
\end{table}
~\\

\subsection{Simplification}
\label{sec:simplification}
At this point, the grammar is operational.
Yet, to enhance its readability, we iteratively apply two transformations to the grammar in a fixed-point iteration:

\begin{description}
\item[Inlining.] We inline non-terminal definitions with a single rule in the referencing context, resulting in a smaller and more readable grammar.
\item[Optional generalization.]  If there are two rules that would be identical after adding a single non-terminal symbol at an arbitrary position in one of the two rules, we merge them. As an example, consider the initial and resulting grammar in~\Cref{fig:optgeneralization}, concisely modeling optional whitespaces.
\end{description}

\begin{figure}[h]
  \centering
  \begin{minipage}[t]{0.5\textwidth}
    \centering
    \emph{Initial grammar}
    \small
    \begin{minted}{xml}
      <statement> ::= <expr> <WS> ;
                    | <expr>      ;
    \end{minted}
  \end{minipage}%
  \hfill %
  \begin{minipage}[t]{0.5\textwidth}
    \centering
    \emph{Resulting grammar}
    \small
    \begin{minted}{xml}
        <statement> ::= <expr> <WS>? ;
    \end{minted}
  \end{minipage}
  \caption{Optional non-terminal generalization}
  \label{fig:optgeneralization}
\end{figure}

\subsection{Reducing Overapproximation}
\label{sec:reducing_overapproximation}

When converting traces into a grammar, we duplicate non-terminal symbols of functions based on their call site.
That is, if a function F is called from two call sites across all execution traces, the grammar would contain two definitions \mintinline{xml}{<F>} and \mintinline[escapeinside=!!]{xml}{<F!$'$!>}.
However, this context is somewhat coarse; a function may parse different strings based on interprocedural constraints that have been passed along a long chain of calls.
We differentiate functions only by their immediate call site, prioritizing completeness of grammars by accepting potential overapproximations.

We address this problem in a general way after mining the initial input grammar by refining overapproximating grammar rules using a dynamic technique.
Note that this step is optional and can be omitted if the overapproximation is acceptable.
We start by generating a set of inputs (e.g., 1000) from the mined input grammar using a grammar-based fuzzer.
Next, we run these inputs in the parser under test and classify them based on the return value as passing (syntactically valid) or failing (syntactically invalid).
For each of the failing inputs%
, we try to make it pass by applying a minimal syntactic change.
Consider the failing input \texttt{5 = 1;} of \tinyc.
The input fails because the left-hand side of the assignment is a number.
This assignment is derived from the grammar rule:

\vspace{0.5cm}
\centerline{\mintinline{xml}{<expr> ::= <test> "=" <expr>}}
\vspace{0.5cm}

The derivation tree of the left-hand side of the assignment is:

\vspace{0.5cm}
\centerline{\mintinline{xml}{<test>-<sum>-<term>-<INT>-5}}
\vspace{0.5cm}

To fix this input --- and the grammar --- automatically, we perform a post-order traversal of the derivation tree (with regards to the initially mined grammar) of the failing input, and repeatedly %
replace each inner node with a randomly generated subtree.
If, after transplanting this subtree, the input now is considered syntactically valid by the parser under test, we have found a refinement candidate.
Let us assume the refinement candidate \texttt{a~=~1;} was generated.
The derivation of the left-hand side is:

\vspace{0.5cm}
\centerline{\mintinline{xml}{<test>-<sum>-<term>-<ID>-a}}
\vspace{0.5cm}

Next, we try to fix the grammar in the most general way, capturing the essence of what differentiates the valid from the invalid input.
We do so by restricting the rule alternatives of the non-terminal symbol corresponding to the root of the replaced subtree with the rule chosen in it.
In this example, the fix would be to update the definition of \mintinline{xml}{<term>} in the grammar:

\vspace{0.5cm}
\centerline{\mintinline{xml}{<term> ::= <ID>}}
\vspace{0.5cm}

To validate this fix, we check whether it causes \textit{underapproximation}.
We parse a set of pre-generated syntactically valid inputs with the updated grammar.
If all these inputs can be parsed, we update the grammar permanently and move on to the next syntactically invalid input, if any.

Otherwise (which is the case here, as no \mintinline{xml}{<INT>} could be generated anymore regardless of the context) we discard the grammar update.
Next, we propagate the rule of the root of the current subtree to its parent node, in order to check if applying the update in the parent context results in a non-underapproximating grammar.
In the example, the next two steps are to refine the rule alternatives of \mintinline{xml}{<sum>}, and \mintinline{xml}{<test>}.
However, both refinements also lead to underapproximating grammars.
Finally, after going up one more level, refining the rule alternatives of \mintinline{xml}{<expr>} does not cause underapproximation:

\vspace{0.5cm}
\centerline{\mintinline{xml}{<expr> ::= <ID> "=" <expr> | <test>}}
\vspace{0.5cm}

In general, in case no non-underapproximating grammar update can be found, we continue with the next failing input.
Given a sufficiently large and diverse set of samples generated during grammar refinement, the result is a refined grammar, which is at least as precise as the initially mined grammar, and has the same recall.

\section{Implementation}
\label{sec:implementation}

Our implementation is based on the symbolic execution engine \klee~\cite{cadar2008klee} which operates on \llvm IR~\cite{DBLP:conf/cgo/LattnerA04}.
We extended \klee both with static analyses as well as dynamic monitoring code.
In total, we added 1190 lines of C++ code to \klee.
On the one hand, we use compiler passes provided by the \llvm compiler infrastructure before starting symbolic execution:
\begin{itemize}
\item We run the \emph{UnifyLoopExits}~\cite{llvmunifyloopexitspass} transformation pass of \llvm, which ensures that loops have a single exit basic block, facilitating our subsequent analysis.
\item We use the \emph{LoopInfoWrapperPass}~\cite{llvmloopinfowrapperpass} to analyze loops in the program and retrieve their header, exiting, and exit basic blocks.
\item We specify the \emph{--switch-type=simple} option to \klee in order to lower switch instructions to branch instructions.
\end{itemize}
On the other hand, we modified \klee to monitor the symbolic execution in order to
\begin{enumerate*}[label=(\arabic*)]
  \item log input accesses;
  \item limit loop iterations and recursive calls;
  \item output all possible solutions for each input byte; and
  \item enable composite analysis of parsers with a dedicated tokenization function.
\end{enumerate*}
 Moreover, we implemented 1210 lines of Python code to infer the input grammar from execution traces output by \klee.

\section{Evaluation}
\label{sec:evaluation}

\APPROACH is the first method capable of inferring input grammars exclusively from the code of parsers.
In contrast, all existing grammar mining techniques~\cite{gopinath2020mimid,kulkarni2021arvada,arefin2024fast,DBLP:conf/kbse/Li0X0C0H24,hoeschele2016autogram} require an initial set of sample inputs to guide the learning process.
Consequently, if no sample inputs are available, these prior approaches are unable to learn an expressive grammar.
Therefore, we do not compare against them in our evaluation.

We conduct a comprehensive evaluation of \APPROACH, addressing the research questions:
\begin{description}
\item[RQ1:] \textbf{Grammar Accuracy.} What are the precision and recall of the mined grammars?
\item[RQ2:] \textbf{Grammar Readability.} How concise are the mined grammars?
\item[RQ3:] \textbf{Scaffolding Effort.} How much manual effort is needed to set up parsers with a lexing stage?%
\item[RQ4:] \textbf{Time Consumption.} How long does it take to mine the grammars?
\item[RQ5:] \textbf{Bugs Found.} Do discrepancies exist between parser implementations and their corresponding language standards?
\end{description}

\subsubsection*{Subjects}

We evaluate \APPROACH on a set of parsers that cover a wide range of languages, summarized in \Cref{tab:evaluation-subjects}.

\begin{table}[h]
  \centering
  \caption{Evaluation subjects}
  \label{tab:evaluation-subjects}
  \rowcolors{2}{white}{gray!20}
  \centering
  \begin{tabularx}{\textwidth}{l l l l l}
      \rowcolor{gray!20}
      \textbf{Subject} & \textbf{Source} & \textbf{Description} & \textbf{Chomsky} & \textbf{Programming}\\
      \textbf{} & \textbf{} & \textbf{} & \textbf{Hierarchy} & \textbf{Language}\\
      \texttt{TINY-C}
      & \cite{subject/tinyc}
      & \emph{Parser for a subset of C.}
      & context-free
      & C
      \\

      \texttt{HARRYDC-JSON}
      & \cite{subject/json}
      & \emph{Parser for JSON.}
      & context-free
      & C
      \\

      \texttt{CJSON}
      & \cite{subject/cjson}
      & \emph{Parser for JSON.}
      & context-free
      & C
      \\

      \texttt{PARSON}
      & \cite{subject/parson}
      & \emph{Parser for JSON.}
      & context-free
      & C
      \\

      \texttt{MJS}
      & \cite{subject/mjs}
      & \emph{Parser for a subset of JavaScript.}
      & context-free
      & C
      \\

      \texttt{LISP}
      & \cite{subject/lisp}
      & \emph{Parser for LISP.}
      & context-free
      & C
      \\

      \texttt{CALC}
      & \cite{subject/calc}
      & \emph{Parser for arithmetic expressions.}
      & context-free
      & C
      \\

      \texttt{CALCCPP}
      & \cite{subject/calccpp}
      & \emph{Parser for arithmetic expressions.}
      & context-free
      & C++
      \\

      \texttt{CGI-DECODE}
      & \cite{subject/cgidecode}
      & \emph{Decoder for CGI.}
      & regular
      & C
      \\

  \end{tabularx}
\end{table}

\subsection{RQ1: Grammar Accuracy}
\label{sec:rq1}

Precision and recall are the core metrics to assess the quality of an inferred grammar.
Intuitively, a precision of 100\% implies that all inputs that were generated from the mined grammar are accepted by the parser under test, i.e., there are no false positives.
In contrast, a recall of 100\% means that all inputs generated from the language we want to learn can be parsed by the mined grammar, i.e., there are no false negatives.
Consequently, computing recall requires a \emph{golden grammar} that serves as a ground truth for the language we want to learn, as we cannot directly generate inputs from the parser under test.
Both metrics are combined in the F1 score, which is the harmonic mean between precision and recall.
These are all established metrics in the field of input grammar mining~\cite{gopinath2020mimid,kulkarni2021arvada,arefin2024fast,DBLP:conf/kbse/Li0X0C0H24}. %
Next, we provide a detailed explanation of how we compute these evaluation metrics, followed by the presentation of the results and a discussion.

To calculate precision, we proceed as follows:
\begin{enumerate}
\item We generate 1000 unique inputs from the inferred grammar using a grammar fuzzer~\cite{fuzzingbook2023:GrammarFuzzer}.
\item We check how many of these inputs are accepted by the parser under test as valid inputs.
\item We output the precision as the percentage of valid inputs among the 1000 total inputs.
\end{enumerate}

As a prerequisite for calculating recall, we define a \emph{golden grammar} for each parser under test, which serves as the ground truth for the language we aim to learn.
Specifically, as we share the evaluation subjects \harrydcjson, \mjs, \tinyc and \cgidecode with \Mimid~\cite{gopinath2020mimid}, we reuse the golden grammars provided by the reproducibility package of \Mimid.
Likewise, for \cjson and \parson, we reuse the golden grammar of the \harrydcjson subject.
We carefully constructed the golden grammar used for \calc and \calccpp by hand. %
For \lisp, we took the golden grammar from the ANTLRv4 git repository~\cite{lisp-golden-grammar}.
To compute recall, we apply the following steps:
\begin{enumerate}
\item We generate 1000 unique inputs from the golden grammar using a grammar fuzzer~\cite{fuzzingbook2023:GrammarFuzzer}, retaining only the inputs that are accepted by the parser under test.
\item We check how many of these inputs can be parsed by the inferred grammar.
\item We output the recall as the percentage of parsable inputs among the 1000 total inputs.
\end{enumerate}

\def\AverageRefinedPrecision{94.5\%\xspace}
\def\AverageRefinedRecall{96.2\%\xspace}
\def\LowestInitialPrecision{15.7\%\xspace}
\def\HighestInitialPrecision{100.0\%\xspace}
\def\LowestRefinedPrecision{65.0\%\xspace}

\begin{table*}[h]
\centering
\begin{threeparttable}
\caption{Grammar accuracy}
\label{tab:accuracy}
\rowcolors{2}{white}{gray!20}
\centering
\begin{tabular}{l|rrr|rrr}
\rowcolor{gray!20}
        & \multicolumn{3}{c}{\textbf{Initial}}
        & \multicolumn{3}{c}{\textbf{Refined}}
        \\
\rowcolor{gray!20}
\textbf{Subject} & \textbf{precision} & \textbf{recall} & \textbf{F1}
                      & \textbf{precision} & \textbf{recall} & \textbf{F1}
                      \\
\href{http://www.iro.umontreal.ca/~felipe/IFT2030-Automne2002/Complements/tinyc.c}{TINY-C}  & 15.7\%    & 100.0\%    & 27.1\%                                                             & 99.6\%   & 100.0\%     & 99.8\%                                                             \\
\href{https://github.com/mystor/simple-lisp-parser-in-c/blob/master/parse.c}{LISP}  & 100.0\%    & 100.0\%    & 100.0\%                                                             & 100.0\%   & 100.0\%     & 100.0\%                                                             \\
\href{https://github.com/cesanta/mjs/blob/master/mjs.c}{MJS}  & 52.2\%    & 100.0\%    & 68.6\%                                                             & 65.0\%   & 84.4\%     & 73.4\%                                                             \\
\href{https://github.com/vrthra/mimid/blob/master/Cmimid/examples/json.c}{HARRYDC-JSON}  & 88.4\%    & 100.0\%    & 93.8\%                                                             & 100.0\%   & 100.0\%     & 100.0\%                                                             \\
\href{https://github.com/DaveGamble/cJSON/blob/master/cJSON.c}{CJSON}  & 87.9\%    & 100.0\%    & 93.6\%                                                             & 99.2\%   & 100.0\%     & 99.6\%                                                             \\
\href{https://github.com/kgabis/parson/blob/master/parson.c}{PARSON}  & 76.8\%    & 78.5\%    & 77.6\%                                                             & 87.1\%   & 81.2\%     & 84.0\%                                                             \\
\href{https://github.com/fbuihuu/parser/blob/master/rdp.c}{CALC}  & 23.3\%    & 100.0\%    & 37.8\%                                                             & 100.0\%   & 100.0\%     & 100.0\%                                                             \\
\href{https://github.com/SaturnMatt/SimpleArithmeticParser/blob/main/Parser/parser.cpp}{CALCCPP}  & 100.0\%    & 100.0\%    & 100.0\%                                                             & 100.0\%   & 100.0\%     & 100.0\%                                                             \\
\href{https://github.com/vrthra/Cmimid/blob/master/examples/cgi_decode.c}{CGI-DECODE}  & 99.9\%    & 100.0\%    & 99.9\%                                                             & 99.9\%   & 100.0\%     & 99.9\%                                                             \\
\end{tabular}

\end{threeparttable}
\end{table*}

\paragraph{\textbf{Discussion.}}
The results of this measurement are shown in \Cref{tab:accuracy}.
The initially mined grammars consistently have a recall of 100\% with the exception of \parson, which reaches a recall of 78.5\%.
The precision of initially mined grammars ranges from \LowestInitialPrecision (\emph{\tinyc}) to 100\% (\emph{\calccpp},\emph{\lisp}).
After the dynamic grammar refinement step, the precision of six out of nine subjects rises, while three subjects remain at a precision of over 99.9\%.
For seven out of nine subjects, precision and recall are over 99\% after refinement, indicating that the mined grammars are highly accurate.

For two subjects, \mjs and \parson, we observe non-optimal results.
\mjs is the most complex parser in our evaluation, consisting of more than 30 parse functions and over 16.000 lines of code. %
Because of this large code size, we cannot rule out that \mjs does not violate some of our assumptions (\Cref{sec:assumptions}).
Moreover, the drop in recall to 84.4\% for \mjs indicates that the grammar refinement step incorrectly accepted an \emph{underapproximating} grammar update.
This could have been avoided by generating a larger set of valid inputs from the initially mined grammar during this step.

For \parson, we investigated the root cause and identified that the code violates our assumption of incremental processing.
Specifically, in function \texttt{is\_decimal}, a loop iterates over a substring of the input in reverse direction, which is not supported by our approach.

Although the results for  \mjs and \parson are not perfect, we argue that they are still valuable and effective for fuzzing applications.
Our results demonstrate that \APPROACH can learn input languages even from highly complex parsers without requiring any sample inputs.

\conclusion{\APPROACH grammars are highly accurate,\\with seven out of nine subjects reaching a precision and recall of over 99\%.}

\subsection{RQ2: Grammar Readability}
\label{sec:rq2}

In addition to being useful for testing, the ability of \APPROACH to infer an input grammar may serve as a reverse engineering aid.
As such, readability and conciseness are crucial for engineers in order to understand and work with the input grammar.
We stress that readability is not merely a quantitative metric.
For instance, the function names \APPROACH extracts from the parser under test to label grammar non-terminals contribute to readability, and are, e.g., not available to black-box approaches which simply label non-terminals with an increasing integer value.
However, consistent with recent related work~\cite{arefin2024fast,DBLP:conf/kbse/Li0X0C0H24}, we also measure quantitative grammar size metrics, shown in \Cref{tab:readability}.
The measured quantities are the number of unique non-terminal symbols (\emph{NT}), the number of rule alternatives (\emph{RA}), the average rule length \emph{l(RA)}, and the sum of all rule lengths \emph{S}.

\begin{table*}[h]
\centering
\begin{threeparttable}
\caption{Grammar readability statistics}
\label{tab:readability}
\rowcolors{2}{white}{gray!20}
\centering
\begin{tabular}{l|rrrr|rrrr}
\rowcolor{gray!20}
        & \multicolumn{4}{c}{\textbf{Initial}}
        & \multicolumn{4}{c}{\textbf{Refined}}
        \\
\rowcolor{gray!20}
\textbf{Subject} & \textbf{NT\tnote{1}} & \textbf{RA\tnote{2}} & \textbf{l(RA)\tnote{3}}  & \textbf{S\tnote{4}}
                      & \textbf{NT\tnote{1}} & \textbf{RA\tnote{2}} & \textbf{l(RA)\tnote{3}}  & \textbf{S\tnote{4}}
                      \\
\href{http://www.iro.umontreal.ca/~felipe/IFT2030-Automne2002/Complements/tinyc.c}{TINY-C}              & 32\    & 119\    & 3.39\  & 403\             & 43\    & 136\    & 3.28\  & 446\             \\
\href{https://github.com/mystor/simple-lisp-parser-in-c/blob/master/parse.c}{LISP}              & 14\    & 287\    & 1.07\  & 306\             & 14\    & 287\    & 1.07\  & 306\             \\
\href{https://github.com/cesanta/mjs/blob/master/mjs.c}{MJS}              & 74\    & 856\    & 1.9\  & 1627\             & 74\    & 473\    & 1.8\  & 852\             \\
\href{https://github.com/vrthra/mimid/blob/master/Cmimid/examples/json.c}{HARRYDC-JSON}              & 51\    & 663\    & 1.15\  & 765\             & 41\    & 389\    & 1.26\  & 491\             \\
\href{https://github.com/DaveGamble/cJSON/blob/master/cJSON.c}{CJSON}              & 45\    & 687\    & 1.28\  & 880\             & 50\    & 691\    & 1.28\  & 887\             \\
\href{https://github.com/kgabis/parson/blob/master/parson.c}{PARSON}              & 48\    & 2212\    & 1.15\  & 2540\             & 58\    & 2459\    & 1.29\  & 3182\             \\
\href{https://github.com/fbuihuu/parser/blob/master/rdp.c}{CALC}              & 22\    & 54\    & 1.65\  & 89\             & 21\    & 50\    & 1.6\  & 80\             \\
\href{https://github.com/SaturnMatt/SimpleArithmeticParser/blob/main/Parser/parser.cpp}{CALCCPP}              & 32\    & 82\    & 2.01\  & 165\             & 32\    & 82\    & 2.01\  & 165\             \\
\href{https://github.com/vrthra/Cmimid/blob/master/examples/cgi_decode.c}{CGI-DECODE}              & 5\    & 281\    & 1.01\  & 283\             & 5\    & 281\    & 1.01\  & 283\             \\
\end{tabular}

\textsuperscript{1}NT = unique non-terminal symbols, \textsuperscript{2}RA = rule alternatives, \textsuperscript{3}l(RA) = average rule length, \textsuperscript{4}S = sum of rule lengths

\end{threeparttable}
\end{table*}

\paragraph{\textbf{Discussion.}}
In general, we observe that the grammar size may increase or decrease after the dynamic grammar refinement step.
The grammar may grow if the definition of a non-terminal symbol is refined only in a specific context, leaving the original definition intact in a different context.
In contrast, the grammar size may also decline if the grammar is refined by deleting rule alternatives.

Note that the high numbers of rule alternatives is mostly due to lexical definitions.
For instance, the character set \verb|[a-z]| would add 26 rule alternatives, whereas \verb|[\x01-\xff]| would even add 255 rule alternatives.
Moreover, parsers may restrict character sets in certain contexts.
As an example, a parser that parses quoted strings as part of its input language (e.g., \json), may not allow the character \verb|"| inside quoted strings, again adding a non-terminal definition with 254 rule alternatives.
We leave it to future work to post-process mined grammars in order to reduce the number of rule alternatives in such cases.
For \parson, we found that the excessive number of rule alternatives is due to the violation of our assumption of incremental processing, as discussed in the previous research question, which leads to a large enumeration of possible characters in a specific context.

\conclusion{Grammars mined by \APPROACH are reasonably sized and readable due to the use of function names as non-terminal symbols.}

\subsection{RQ3: Scaffolding Effort}
\label{sec:rq3}

The effort required to prepare a parser for analysis with \APPROACH is roughly equivalent to that of conventional software testing, i.e., setting up the program such that it accepts input from a test generator.

For subjects with a lexing stage (\mjs, \tinyc, \lisp), \APPROACH requires \emph{glue code} that specifies the information flow between lexer and parser; see \Cref{fig:harness-tokenization-function} and \Cref{fig:proxy-tokenization-function} for \tinyc.
To quantify the manual effort to prepare these subjects, \Cref{tab:manual-effort} summarizes the number of unique lines of code we implemented.
We expect the manual effort for this stage to take less than ten minutes.

\begin{table}[h]
\caption{Manual effort (unique lines of code)}
\label{tab:manual-effort}
\rowcolors{2}{gray!20}{white}
\begin{tabular}{lrr}
\rowcolor{gray!20}
\textbf{Subject} & \textbf{\#L Tokenization Harness}           & \textbf{\#L Tokenization Proxy} \\
\href{http://www.iro.umontreal.ca/~felipe/IFT2030-Automne2002/Complements/tinyc.c}{TINY-C} & 3 & 2 \\
\href{https://github.com/mystor/simple-lisp-parser-in-c/blob/master/parse.c}{LISP} & 5 & 5 \\
\href{https://github.com/cesanta/mjs/blob/master/mjs.c}{MJS} & 5 & 10 \\
\end{tabular}
\end{table}

In principle, an industrial implementation could eliminate this manual effort by automatically detecting information flow between lexer and parser and matching it against a number of common patterns.
However, this is beyond the scope of this work.

\conclusion{Scaffolding code for the evaluation subjects with a lexing stage \\ consists of between 5 and 15 unique lines of code.}

\subsection{RQ4: Time Consumption}
\label{sec:rq4}

In this research question, we investigate the time requirements of \APPROACH and analyze its distribution over the different analysis steps.
We ran \APPROACH on each evaluation subject with a memory limit of 16 GB and a time limit of 24 hours for parser analysis, and two hours for tokenization analysis for parsers with a lexing stage.

\begin{table}[t]
\caption{Time consumption}
\label{tab:time}
\rowcolors{2}{white}{gray!20}
\centering
\begin{tabular}{l|rrr|r}

\rowcolor{gray!20}
                      & \textbf{Lexer}    & \textbf{Parser}   & \textbf{Grammar}    & \textbf{Total} \\
\rowcolor{gray!20}
\textbf{Subject} & \textbf{Analysis} & \textbf{Analysis} & \textbf{Refinement} & \textbf{Time} \\
\href{http://www.iro.umontreal.ca/~felipe/IFT2030-Automne2002/Complements/tinyc.c}{TINY-C}  & 00h:09m  & 05h:06m  & 00h:03m & 05h:18m \\
\href{https://github.com/mystor/simple-lisp-parser-in-c/blob/master/parse.c}{LISP}  & 00h:02m  & 24h:15m  & \textless00h:01m & 24h:18m \\
\href{https://github.com/cesanta/mjs/blob/master/mjs.c}{MJS}  & 00h:43m  & 09h:19m  & 01h:16m & 11h:19m \\
\href{https://github.com/vrthra/mimid/blob/master/Cmimid/examples/json.c}{HARRYDC-JSON}  & N/A  & 11h:32m  & \textless00h:01m & 11h:32m \\
\href{https://github.com/DaveGamble/cJSON/blob/master/cJSON.c}{CJSON}  & N/A  & 24h:27m  & 00h:01m & 24h:28m \\
\href{https://github.com/kgabis/parson/blob/master/parson.c}{PARSON}  & N/A  & 12h:36m  & 00h:28m & 13h:04m \\
\href{https://github.com/fbuihuu/parser/blob/master/rdp.c}{CALC}  & N/A  & 25h:18m  & 00h:01m & 25h:20m \\
\href{https://github.com/SaturnMatt/SimpleArithmeticParser/blob/main/Parser/parser.cpp}{CALCCPP}  & N/A  & 24h:08m  & \textless00h:01m & 24h:08m \\
\href{https://github.com/vrthra/Cmimid/blob/master/examples/cgi_decode.c}{CGI-DECODE}  & N/A  & \textless00h:01m  & \textless00h:01m & \textless00h:01m \\
\end{tabular}
\end{table}

\Cref{tab:time} summarizes the time consumption.
Column \emph{Lexer Analysis} shows the time spent on analyzing the lexer of parsers with a lexing stage.
Column \emph{Parser Analysis} indicates the time required by the parser analysis (including converting traces to the grammar), whereas column \emph{Grammar Refinement} lists the times spent on refining the grammars.

\conclusion{\APPROACH takes between one minute and 25 hours to infer grammars,\\requiring less than 16 GB of memory.}

\subsection{RQ5: Bugs Found}
\label{sec:rq5}

From a developer perspective, it is crucial to ensure that the parser implementation matches the language specification.
In this research question, we evaluate if the implemented parsers are \emph{too permissive}, accepting some inputs that are not accepted by the specification.
We assume the specification is provided by the golden grammar used in RQ1.
As an example, we compare the grammar inferred from the \harrydcjson parser with the golden \json grammar to see if it accepts more than the golden grammar allows.

We address RQ5 by
\begin{enumerate}[label=(\arabic*)]
\item generating inputs from the mined grammar,
\item retaining those inputs that can be parsed by the parser implementation under test, and
\item checking whether the golden grammar can also parse these inputs.
\end{enumerate}
In case the golden grammar cannot parse an input that is accepted by the implementation, we say that the implementation is too permissive.
We applied this methodology on the evaluation subjects, and present some of the mismatches we discovered.

\begin{description}
\item[JSON.]
For the \harrydcjson parser, we detected a syntactic mismatch in the grammar extracted by \APPROACH.
   In \Cref{fig:json-grammar}, consider the definition of
  \mintinline[escapeinside=!!]{xml}{<json_parse_object!$'$!>}.
  We see that keys of objects may be any \mintinline{xml}{<json_parse_value>}, including strings, numbers, arrays, and objects---say, \texttt{\char123 1: "foo"\char125}.
  However, the \json specification\footnote{\url{https://ecma-international.org/publications-and-standards/standards/ecma-404/}} and our golden grammar only permit strings as keys of objects.
The issue was validated with concrete inputs using the \harrydcjson parser as oracle.
It is easy to pinpoint such issues in \APPROACH grammars.

\item[\mjs.]
For \mjs, we detected two syntactic mismatches, which are not allowed in JavaScript.
\Cref{fig:mjs-let} shows an excerpt of the mined \mjs grammar, defining variable declarations.
\begin{enumerate*}
  \item The \mintinline{xml}{<opt_comma>} definition shows that in \mjs \texttt{let} constructs, commas are \emph{optional} between variable names.
As a consequence, it is possible to declare multiple variables by separating them with a whitespace only (e.g., \texttt{"let a b"}).
\item
Moreover, variable declarations may contain a trailing comma.
For instance, \texttt{"let a, b,"} is valid in \mjs.
\end{enumerate*}
\end{description}

\begin{figure}[h]
  \begin{center}
    \begin{minipage}{.6\textwidth}
    \small\linespread{1.1}
      \begin{minted}{xml}
<statement> ::= "let" <let_loop> | ...
<let_loop>  ::= <ascii_str> <opt_comma> <let_loop> | ...
<opt_comma> ::= '' | ','
      \end{minted}
    \end{minipage}
  \end{center}
  \caption{\mjs \texttt{let} declarations mined by \APPROACH (simplified)}
  \label{fig:mjs-let}
\end{figure}

These examples highlight one of the advantages of symbolic input grammar mining.
Sample-based grammar miners would typically learn from \emph{well-formed} inputs that do not include faulty input features.
Hence, they would be unlikely to find potential bugs associated with such borderline valid inputs.
Developers can use \APPROACH to catch mismatches in their implementation easily.
The alternative would be to manually assess the parser implementation for syntactic mismatches, which is time-consuming and error-prone.

\conclusion{\APPROACH can learn grammars accurately, capturing even subtle details.}

\section{Related Work}
\label{sec:related_work}

Automated inference of input grammars has been studied extensively, with the vast majority of existing work focusing on learning input grammars from examples.

\subsection{Language Inference}

In her seminal paper~\cite{DBLP:journals/iandc/Angluin87}, Angluin presented the L* algorithm, which can learn regular languages by means of active learning.
This algorithm requires a \emph{teacher} that serves as a parse oracle and can decide whether the \emph{learned language} is equivalent to the target language, providing a counterexample in case of a negative answer.
In the same paper, Angluin also presented the L\textsuperscript{cf} algorithm to address learning of context-free languages in a similar way.
However, this algorithm is not practical due to its strong assumptions on the grammar and information required by the learner.
In general, black-box inference of context-free grammars was shown to be computationally infeasible by Angluin~et~al.~\cite{DBLP:conf/stoc/AngluinK91}.

Therefore, recent black-box approaches rely on the fact that context-free languages might not be as complicated in practice~\cite{kulkarni2021arvada}, giving raise to heuristics-infused black-box learning.
Moreover, white-box approaches~\cite{gopinath2020mimid, hoeschele2016autogram, schroeder2022} leverage knowledge of the program code to draw conclusions about the input language accepted by the parser.

\subsection{Black-Box Grammar Inference}

\Arvada~\cite{kulkarni2021arvada} is a black-box approach to learn context-free grammars based on a set of positive sample inputs and a parse oracle.
It starts by generating \emph{flat} parse trees (i.e., the root node has one child for each input character) for each input.
Next, it gradually adds \emph{structure} to these trees by combining sibling nodes under a new intermediate parent node (\emph{bubbling} operation), retaining only those bubbles that can be merged later.
The validity of a merge operation is checked using the parse oracle.
As a black-box approach, \Arvada is agnostic of the implementation.
However, it is unlikely to synthesize features not found in sample inputs.

\TreeVada~\cite{arefin2024fast} finds that the \Arvada approach is non-deterministic, and requires very short sample inputs to be available. 
\TreeVada introduces a pre-structuring heuristic, which avoids any generalization attempts that would result in unbalanced brackets.

\Kedavra~\cite{DBLP:conf/kbse/Li0X0C0H24} is an improvement over TreeVada and follows a similar technique.
It first decomposes sample inputs on token boundaries and then infers the grammar incrementally, resulting in a more efficient and effective inference process.

By construction, all these approaches require sample inputs on top of the program under test---in contrast to \APPROACH, which only needs the input-processing program, but also the ability to analyze it.
Given a sufficiently large and sufficiently diverse of inputs to learn from, \Arvada, \TreeVada, and \Kedavra can all produce useful grammars.
However, if such a set of inputs is already available, one may also use it directly for tasks such as testing.

\subsection{White-Box Grammar Inference}

\Autogram~\cite{hoeschele2016autogram} is the seminal work on white-box mining of context-free input grammars.
This approach is based on dynamically tracking data flow in a parser when it processes the input.
Hence, \Autogram depends on sample inputs and requires that data flow is present in the first place.

Being the approach most closely related to ours, \Mimid~\cite{gopinath2020mimid} is a white-box approach to learn context-free grammars from recursive descent parsers. However, in contrast to \APPROACH, \Mimid requires sample inputs.
In essence, \Mimid tracks how the parser implementation processes inputs, labeling an input character with the execution context (based on the call stack and control-flow structures) at the point of its acceptance.
This labeling allows to derive grammar-inducing parse trees from the seed inputs, which are further generalized using experiments.

All these approaches are \emph{dynamic} and require sample inputs---in contrast to \APPROACH, which only needs the input-processing program.
And again, if a feature is not present in the set of sample inputs, it is unlikely to be learned by these approaches.

\subsection{Static Grammar Inference}

The only other work we are aware of that attempts to statically mine grammars from code comes from Schröder et al.~\cite{schroeder2022}.
Like \APPROACH, this approach does not require sample inputs.
It focuses on \emph{short ad-hoc parsers} implemented in Python, such as \texttt{xs = map(int, s.split(','))}, extracting input grammars using summaries of commonly used string functions.
In contrast, \APPROACH is based on symbolic execution and can handle much larger recursive descent parsers implemented in~C.

The idea of \APPROACH, including first results, were first published in an FSE~2024 short paper~\cite{bettscheider2024look}.

\section{Limitations and Future Work}
\label{sec:limitations_future_work}
While \APPROACH makes significant contributions, acknowledging its limitations opens exciting opportunities for future work.

\begin{description}
\item[Accurate token mining.]
In \APPROACH, we learn tokens by generating token instances using symbolic execution and matching them with predefined patterns.
While we expect most tokens to belong to common categories such as uppercase strings or digits, in general, tokens can represent any regular language, which may not always align with one of the predefined patterns, leading to overapproximation.
Future work may learn tokens more accurately---e.g., using active learning of regular expressions~\cite{DBLP:journals/iandc/Angluin87}.

\item[Automatic harness construction.]
While \APPROACH currently only requires a small amount of manual work to handle parsers with a lexing stage, we will investigate ways to further automate our analysis for such parsers.

\item[Semantic constraints.]
Our \APPROACH approach is entirely focused on learning the syntax of inputs.
Input formats may additionally have non-syntactic constraints, such as \emph{a variable must be defined before it used}.
As part of our ongoing research, we explore ways to \emph{extract} such semantic constraints from a given program in order to \emph{lift} it to the input specification, thereby enhancing its precision even further.
The input specification language ISLa~\cite{steinhoefel2022isla} allows to specify such semantic constraints in conjunction with context-free input grammars.

\item[Other domains.]
In this work, we have investigated the inference of context-free input grammars from recursive descent parsers.
Such parsers typically process recursive text-based languages.
In the future, we would like to extend our approach to other domains, such as text-based and binary protocols, as well as parsers processing binary data.

\item[Assumptions.] Finally, the assumptions in \Cref{sec:assumptions} can all be read as limitations, too.
\end{description}

\section{Conclusion}
\label{sec:conclusion}

Accurate input specifications allow fuzzers to efficiently and comprehensively test programs.
Up until now, all approaches to automatically infer input grammars have required a set of \emph{sample inputs} as a starting point.
Consequently, these approaches could only learn input features that were present in the provided sample inputs, making them inherently dependent on the quality and diversity of the sample inputs.
In this paper, we present \APPROACH, the first static input grammar mining technique, which produces accurate input grammars entirely without the need for sample inputs.
Our evaluation shows that grammars produced by \APPROACH can have near-perfect precision and recall, and are very accurate at capturing the input format accepted by the program under test.

\section{Data Availability}
\label{sec:data-availability}

All of \APPROACH and experimental data is available at:
\begin{center}
\url{https://github.com/leonbett/stalagmite}
\end{center}

\begin{acks}
This work was funded by the German Federal Ministry of Education and Research (BMBF, project CPSec - 16KIS1564K).
\end{acks}

\bibliographystyle{ACM-Reference-Format}
\bibliography{main}

\end{document}